**Twist-induced altermagnetism in a metallic van der Waals antiferromagnet**

*Alberto M. Ruiz, Andrei Shumilin, Rafael González-Hernández, José J. Baldoví[*]*

A.M. Ruiz, A. Shumilin, J. J. Baldoví

Instituto de Ciencia Molecular, Universitat de València, Catedrático José Beltrán 2, 46980 Paterna, Spain.

R.González-Hernández

Departamento de Física y Geociencias, Universidad del Norte, Barranquilla, Colombia.

E-mail: j.jaime.baldovi@uv.es

**Abstract**

Altermagnetism –a magnetic state characterized by spin-polarized electronic bands at zero net magnetization– offers a promising route for next-generation spintronic devices. In two-dimensional (2D) magnets, twist engineering enables its realization by breaking the combined inversion and time-reversal symmetry ($\mathcal{P}T$) while preserving crystal symmetries that ensure the altermagnetic order. Here, by first-principles calculations and symmetry analysis, we demonstrate that twist engineering applied to the recently synthesized metallic van der Waals antiferromagnet Co-doped bilayer $Fe_3GaTe_2$ ($Fe_2CoGaTe_2$) provides a robust platform for altermagnetism. By twisting two layers of $Fe_2CoGaTe_2$, the $\mathcal{P}T$ symmetry between opposite spin sublattices is broken, resulting in a non-relativistic *i*-wave altermagnetic state with spin-splitting up to 138 meV. In the absence of spin-orbit coupling (SOC) the electronic states are spin-degenerate along six high-symmetry directions while the inclusion of SOC preserves this degeneracy along the three directions protected by twofold rotation axes. Furthermore, we unveil the microscopic mechanisms governing the magnetic behaviour in twisted $Fe_2CoGaTe_2$. Our results establish twist engineering and metallic Fe-based van der Waals antiferromagnets as versatile platforms to realize 2D van der Waals altermagnetism, with potential for designing high-efficiency ultrathin nanodevices.

## 1. Introduction

Altermagnetism is a novel magnetic state characterized by the coexistence of spin-polarized electronic bands and zero net magnetization, distinguishing it from conventional ferromagnets and antiferromagnets.[1–4] This combination offers significant advantages for next-generation spintronic applications, such as the spin Hall effect and magneto-optical Kerr effect, primarily due to the absence of stray fields and the promise of ultrafast spin dynamics. Unlike conventional antiferromagnets, in which opposite spin sublattices are related by translational ($\tau$) or inversion ($\mathcal{P}$) symmetry, the altermagnetic (AM) state requires that these sublattices be connected via specific rotational ($\mathcal{C}$) or mirror ($\mathcal{M}$) symmetry operations. Consequently, the realization of altermagnetism requires the breaking of the combined $\mathcal{P}$ and time-reversal ($T$) symmetries, known as $\mathcal{P}T$, which enables spin splitting while preserving specific $\mathcal{C}$ or $\mathcal{M}$ symmetries that enforce the characteristic alternation of spin polarization in momentum space.[5–8]

Two-dimensional (2D) magnetic materials provide an attractive platform for realizing altermagnetism, as their tunable lattice symmetry and magnetic interactions allow for fine control of their properties. However, many well-studied van der Waals (vdW) A-type antiferromagnets, such as $CrI_3$ and CrSBr, possess $\tau$ or $\mathcal{P}$ symmetries between opposite spin sublattices.[9,10] This enforces conventional antiferromagnetic (AF) behavior and forbids the momentum-space spin splitting required for an AM state.[11–13] Among the existing 2D vdW magnetic materials, the recently synthesized vdW itinerant ferromagnet $Fe_3GaTe_2$ has attracted significant attention due to its above-room-temperature magnetism ($T_C$ = 350−380 K), metallic character and out-of-plane anisotropy.[14–21] Indeed, in the context of altermagnetism, metallic altermagnets are particularly appealing due to the possibility of direct electrical control and efficient manipulation of spin currents via applied electric fields.[22] Nevertheless, altermagnetism is prohibited in $Fe_3GaTe_2$ given that (*i*) it exhibits a FM ground state and (*ii*) spin sublattices of adjacent layers are connected through $\mathcal{P}$ symmetry. Partial substitution of Fe with Co in $(Fe_{1-x}Co_x)_3GaTe_2$ has recently been experimentally demonstrated to induce a transition from ferromagnetic (FM) to AF ground state, solving (*i*), whereas (*ii*) is still present, thus preventing AM spin splitting.[23–25]

As a consequence of the lack of vdW 2D magnetic materials that naturally host AM order, several theoretical strategies have been proposed to induce altermagnetism. These include the application of a gate-induced out-of-plane electric field, ligand substitution or creation of heterostructures.[26–28] Another well-stablished route is the realization of Janus 2D magnetic materials, in which the two faces of the material are chemically different, inducing an out-of-plane asymmetry within the layer.[29–32] These approaches break the combined $\mathcal{PT}$ symmetry while preserving selected symmetry operations relating opposite spin sublattices.

Besides these approaches, twistronics –the tuning of the relative twist angle between layers in vdW heterostructures– has emerged as an unprecedented platform for exploring novel electronic and magnetic states since 2018.[33–35] In the case of altermagnetism, it has been proposed that AM order can be induced upon rotation between two layers.[36–38] However, most materials explored to date exhibit semiconducting behavior and some of these works rely on the assumption that antiferromagnetism present in the parent non-rotated compounds is preserved upon twisting. Consequently, identifying metallic vdW materials that can retain AF interlayer coupling under twisting remains a central challenge for the realization of twist-induced altermagnetism in vdW materials.

Here, we overcome these limitations by demonstrating that metallic Co-doped bilayer $Fe_3GaTe_2$ hosts robust AM order upon twisting. Using first-principles calculations, we show that Co doping drives $Fe_3GaTe_2$ from a FM to an AF state, consistent with recent experimental reports. Importantly, our symmetry analysis reveals that upon twisting, the AF interlayer coupling is preserved, while the rotation breaks the $\mathcal{PT}$ symmetry that connects opposite spin sublattices. This breaking gives rise to a non-relativistic spin-split electronic band structure with a maximum splitting of 138 meV around the Fermi level. Our results position twistronics as a versatile platform for inducing altermagnetism in twisted 2D magnetic materials, as well as the use of $Fe_3GaTe_2$ for efficient spintronic applications.

## 2. Results and Discussion

$Fe_3GaTe_2$ is a vdW magnetic material that crystallizes in a hexagonal layered structure belonging to the centrosymmetric space group $P6_3/mmc$ (N$_0$. 194). Each layer comprises five atomically distinct sublayers: the outermost planes are formed by Te, followed by two Fe sublayers, while the central plane is constituted by Ga and Fe atoms (Figure 1a). Between adjacent $Fe_3GaTe_2$ layers, there is an inversion center and thus the system shows $\mathcal{P}$ symmetry. We first perform first-principles calculations on bilayer $Fe_3GaTe_2$, for which the symmetry is reduced to P-3m1 (N$_0$. 164).[39] The optimized lattice parameters using the GGA scheme are $a = b = 4.03$ Å and the computed averaged magnetic moment for Fe atoms is 2.03 $\mu_B$ (Figure S1 and Table S1). The material exhibits FM coupling between layers, which is determined by computing the interlayer exchange coupling ($J_{int}$). It is defined as $J_{int} = E_{AF} - E_{FM}$ and is found to be 1.18 meV/Fe, correctly capturing the experimentally observed magnetic ground state. This spin alignment gives rise to a Zeeman-type splitting of the electronic bands, as illustrated in Fig. 1a. Furthermore, $Fe_3GaTe_2$ exhibits out-of-plane anisotropy with spins aligned along the *c* axis, quantified by computing the magnetic anisotropy energy (MAE), defined as the energy difference between in-plane and out-of-plane spin orientations (MAE = $E_\parallel - E_\perp$), resulting in MAE = 0.45 meV/Fe.[40] Figure S2 depicts the electronic band structure of bilayer $Fe_3GaTe_2$, showing multiple bands crossing the Fermi level, where electron and hole pockets are observed around the K and Γ points, respectively, in agreement with ARPES measurements and previous theoretical reports.[19,20,41]

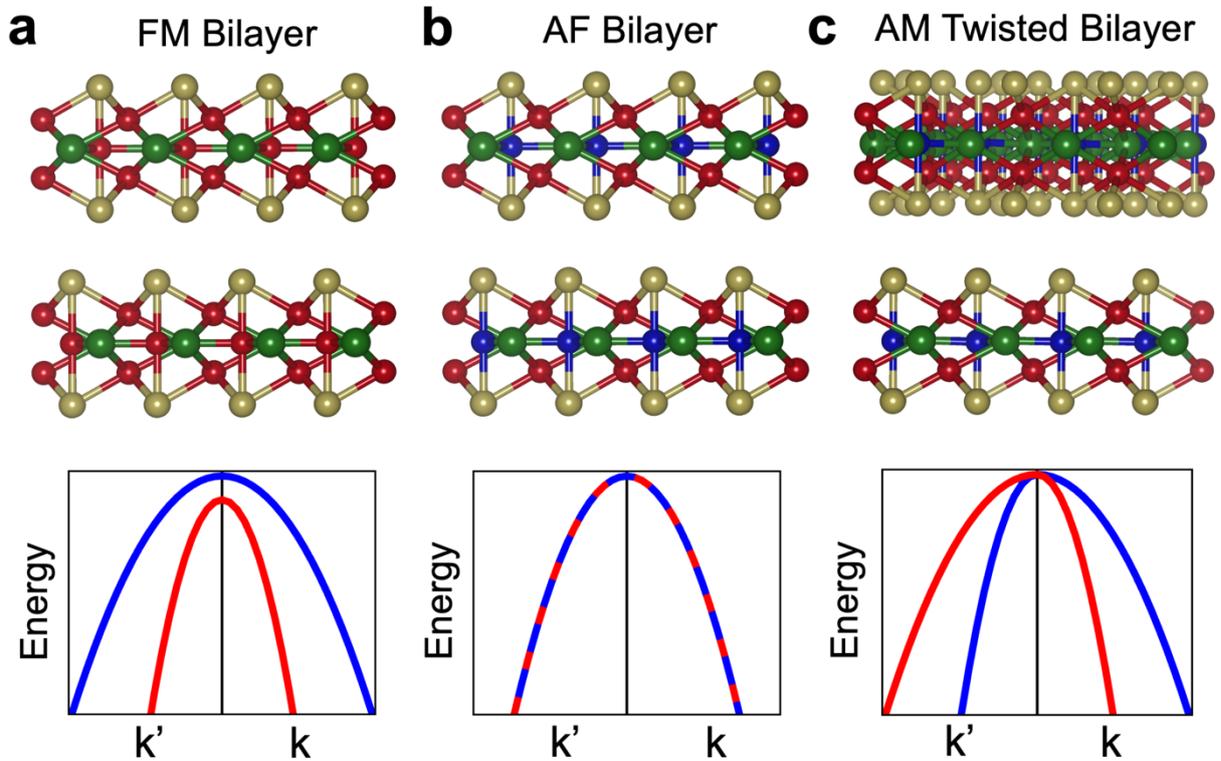

Figure 1. a) Lateral view of bilayer $Fe_3GaTe_2$ with its corresponding schematic FM-like band structure. b) Lateral view of bilayer $Fe_2CoGaTe_2$ and its correspondent schematic degenerate AF band structure. c) Lateral view of 21.79° twisted bilayer $Fe_2CoGaTe_2$, where the non-centrosymmetric structure gives rise to AM spin splitting. Color code: Fe (dark red), Co (dark blue), Ga (green), and Te (yellow). Blue (red) color in the band structure indicates the projected spin up (down) states.

We next consider the substitution of Fe with Co in $(Fe_{1-x}Co_x)_3GaTe_2$, focusing on a doping level of x = 1/3, corresponding to the composition $Fe_2CoGaTe_2$. This concentration is selected based on recent experimental reports showing that Co doping induces interlayer antiferromagnetism in $(Fe_{1-x}Co_x)_3GaTe_2$ for 0.15 < x < 0.40.[23–25] Among the possible substitutional arrangements, the energetically most favorable configuration is the one in which

Co atoms occupy the Fe sites in the central layer (Figure 1b; Figure S3 and Table S2),[42] retaining the centrosymmetric P-3m1 space group.[25] For $Fe_2CoGaTe_2$, the computed magnetic moments are 2.56 $\mu_B$/Fe and 0.88 $\mu_B$/Co and the interlayer exchange coupling becomes strongly AF, where $J_{int}$ = -2.44 meV per magnetic atom. The out of plane anisotropy is preserved, showing an enhanced value of MAE = 1.44 meV per magnetic atom. Spin–space–group (SSG) analysis reveals that $Fe_2CoGaTe_2$ belongs to a type-III AF group that contains the antiunitary generator [-1∥-1], where in the notation [S∥R] the first entry S denotes the spin-space operation (-1 corresponds to spin inversion) and the second entry R specifies the associated real-space operation (-1 is spatial inversion). This [-1∥-1] operation connects the two opposite Fe (or Co) spin sublattices and, because it squares to -1, it enforces a Kramers-like double spin degeneracy of energy bands in the full Brillouin zone, resulting in conventional metallic AF band structure (Figure 1b and Figure S4). As in $Fe_3GaTe_2$, we note the presence of electron (around K) and hole (around Γ) pockets. Compared to $Fe_3GaTe_2$, these pockets are shifted downward due to the additional electrons introduced by Co atoms (Figure S4).

Then, we analyse the $Fe_2CoGaTe_2$ twisted bilayer upon a rotation angle of 21.79°. This particular twist angle is selected since it breaks the $\mathcal{P}T$ symmetry while inducing a zero strain in both sublattices,[43] and it gives rise to an AM spin splitting in the non-relativistic limit as represented in Figure 1c. The magnetic moments of Fe and Co atoms remain essentially unchanged relative to the untwisted structure, whereas the interlayer coupling retains its AF character, with $J_{int}$ = -0.54 meV per magnetic atom. The reduced magnitude of $J_{int}$ with respect untwisted $Fe_2CoGaTe_2$ arises from the increased spacing between magnetic centers, which diminishes the orbital overlap mediating the AF exchange interactions. Specifically, the interlayer separation increases from 2.94 Å in untwisted bilayer $Fe_2CoGaTe_2$ to 3.25 Å upon rotation. Nevertheless, the interlayer coupling remains stronger than in other 2D vdW magnets such as CrSBr ($J_{int}$ ≈ -0.1 to -0.2 meV/Cr), and comparable to bilayer $CrPS_4$ ($J_{int}$ ≈ -0.50 meV/Cr).[44–46] Importantly, calculations performed using the LDA functional confirm the robustness of the interlayer AF coupling, yielding $J_{int}$ = -0.86 meV per magnetic atom. Additionally, twisted $Fe_2CoGaTe_2$ shows spins along the out-of-plane *c* axis, where MAE = 1.64 meV/magnetic atom. In twisted-$Fe_2CoGaTe_2$, the relative rotation between layers breaks the global [-1∥-1] ($\mathcal{P}T$ symmetry). As a consequence, the two opposite spin sublattices are no longer related by $\mathcal{P}T$ symmetry, but instead are connected through 180° spatial rotations combined with spin inversions, encoded in the SSG operations: [-1 ∥ $2_{120}$], [-1 ∥ $2_{1-10}$], and [-1 ∥ $2_{210}$]. In addition, collinear magnet exhibits the symmetry $T$[-1 ∥ 1], which combines time-reversal with a spin-flip. Meanwhile, the unitary operator [ 1 || $3_{001}$ ] connects site within the same spin sublattice. These symmetry operations act in the real space and are illustrated in Figure 2a (see also Figure S5 for details). Their combined effect constrains the electronic structure in the reciprocal space, leading to a characteristic spin splitting pattern in which the spin projection alternates every 30º in momentum space (Figure 2b). As a result, spin splitting becomes symmetry-allowed in the nonrelativistic limit, and the corresponding reduction in SSG operations identifies the twisted structure as an *i*-wave altermagnet.

This behaviour is clearly reflected in the electronic band structure (Figure 2c). The material exhibits metallic character, with several bands crossing the Fermi level. Spin-degenerate bands are observed along the Γ−M−K path, whereas spin-splitting emerges along the K−(−)K direction, showing a more pronounced effect in the valence bands than in the conduction bands. Within the studied energy range, the maximum spin-splitting reaches 138 meV for bands located approximately 0.2 eV below the Fermi level, while at the Fermi energy the largest splitting is 70 meV.

The spin-resolved Fermi surface shown in Figure 2d exhibits a set of star-like sheets arranged with sixfold symmetry. The color scale encodes the expectation value of the spin component $\langle S_z(k) \rangle$, with blue and red indicating opposite spin polarizations in the nonrelativistic limit. The Fermi-surface regions are spin-polarized, but their contributions to the total magnetization vanish due to the symmetry-enforced compensation. The Fermi-surface topology exhibits the characteristic AM pattern predicted by the symmetry analysis: the spin polarization alternates every 30°, reflecting the momentum-dependent spin splitting imposed by the underlying SSG operations. In addition, the overall Fermi-surface pattern retains a 120° rotational symmetry, consistent with the [1||$3_{001}$] symmetry of twisted bilayer $Fe_2CoGaTe_2$.

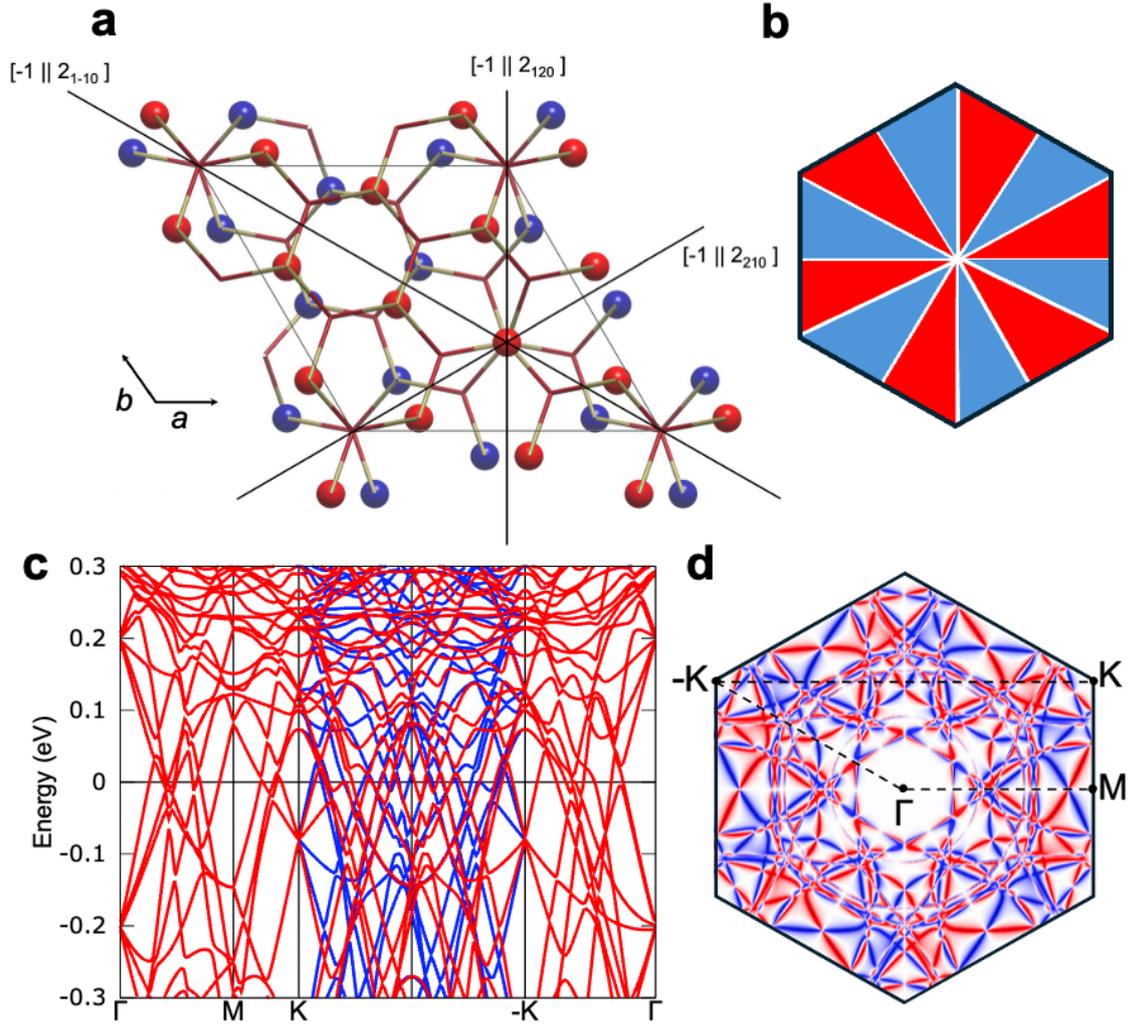

Figure 2. a) Top view of twisted bilayer $Fe_2CoGaTe_2$, where the blue and red balls represent Co atoms from the bottom and top layer, respectively. For clarity, we only represent Co atoms of the material. Additionally, we show the symmetry operations relating opposite Co spin sublattices. b) Schematic visualization of the spin-split bands according to the spin symmetry operations on the first Brillouin zone. c) Non-relativistic electronic band structure of twisted bilayer $Fe_2CoGaTe_2$ along the high-symmetry path $\Gamma-M-K-(-)K-\Gamma$ and its corresponding d) spin-resolved Fermi surface at the Fermi level. The bands with spin up and down projection are shown by blue and red, respectively.

In the presence of SOC, the material belongs to the non-centrosymmetric magnetic space group P312 (#149.21). In this case, the twofold rotations act simultaneously on spatial coordinates and spin directions, without requiring an additional time-reversal operation. This symmetry forbids spin degeneracy perpendicular to the rotation axis, and in particular along the out-of-plane direction (see Figure S6 for details). As a result, the electronic band structure along the directions of the twofold rotation axes remains degenerate, even when SOC is included. Figure 3 shows the electronic states at the Fermi level calculated for the twisted $Fe_2CoGaTe_2$ bilayer without and with SOC, respectively. Consistent with our symmetry analysis, the states are spin-degenerate along all six directions in the absence of SOC. When SOC is included, the spin degeneracy is preserved along the directions of the twofold rotation axes (indicated by green lines), but lifted along the perpendicular directions (black lines), where no symmetry protection is provided by the magnetic space group #149.21.

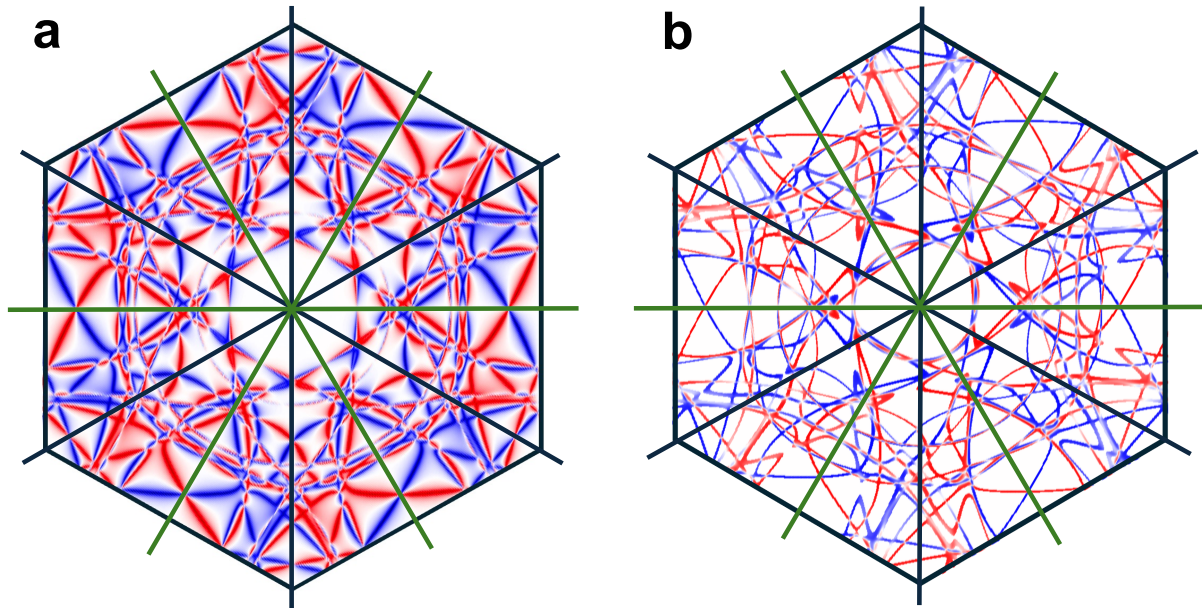

Figure 3. Spin-resolved Fermi surface along with the planes along the directions of the twofold rotation axes (green lines) and perpendicular to them (black lines) for $Fe_2CoGaTe_2$ a) without and b) with SOC effects. Blue and red lines represent the normalized expectation value of the spin component ⟨$S_z$ (k)⟩.

These results show that magnetic anisotropy and spin orientation are essential for correctly describing altermagnetism in twisted bilayers. Due to its out-of-plane anisotropy, $Fe_2CoGaTe_2$ retains zero net magnetization upon twisting, inducing an AM state. Note that 90°-twisted CrSBr, although theoretically proposed as an altermagnet in the non-relativistic limit,[36,47] experimental findings have reported that spins remain aligned along the easy *b*-axis of each layer.[48–50] This results in an orthogonal spin arrangement with a global FM state, precluding altermagnetism.

We also examine whether twisting pristine bilayer $Fe_3GaTe_2$ can induce an AF ground state, and thereby give rise to altermagnetism. However, we find that upon twisting, $Fe_3GaTe_2$ remains FM with an interlayer exchange coupling of $J_{int}$ = 0.61 meV/Fe. The magnitude of $J_{int}$ is reduced compared to the untwisted case ($J_{int}$ = 1.14 meV/Fe) due to the increased interlayer spacing (3.4 Å vs 2.94 Å). Consequently, the electronic bands exhibit conventional FM spin splitting throughout the Brillouin zone (Figure S7).

The exchange interactions in twisted Fe$_2$CoGaTe$_2$ are obtained using a spin Hamiltonian of the form:

$$H = -\sum_{i \neq j} J_{ij}\vec{S}_i \cdot \vec{S}_j$$

where J$_{ij}$ represent the isotropic exchange interactions, S$_i$ and S$_j$ are the magnetic moments of different sites and normalized to 1.

The J$_{ij}$ for each layer can be divided into two main categories. The first comprises inter-plane couplings, which include the interactions between Fe$_1$ and Fe$_2$, denoted as J$_{Fe1–Fe2}$, as well as J$_{Fe1–Co}$ (between Fe$_1$ and Co) and J$_{Fe2–Co}$ (Fe$_2$ and Co). In bulk and monolayer Fe$_2$CoGaTe$_2$, J$_{Fe1–Co}$ and J$_{Fe2–Co}$ are equivalent given that Fe$_1$ and Fe$_2$ share identical local environment. This equivalence is lost in the bilayer, where Fe$_1$ faces vacuum on one side, while Fe$_2$ is adjacent to the vdW gap and the neighbouring Fe$_2$CoGaTe$_2$ layer (Figure 4a). The second category corresponds to in-plane exchange, involving atoms located at the same z coordinate, including J$_{Fe1–Fe1}$, J$_{Fe2-Fe2}$ and J$_{Co–Co}$. As shown in Figure 4b, the selected exchange interactions decrease with increasing distance and vanish at 16 Å. For clarity, Figure 4b displays J$_{Fe1–Fe2}$, J$_{Fe1–Co}$, J$_{Fe1–Fe1}$ and J$_{Co–Co}$, since the corresponding interactions J$_{Fe2–Co}$ and J$_{Fe2–Fe2}$ are almost identical to J$_{Fe1–Co}$ and J$_{Fe1–Fe1}$ (Figure S8), respectively, owing to the nearly identical magnetic moments of Fe$_1$ and Fe$_2$ (2.59 µ$_B$ vs 2.54 µ$_B$). We observe that the inter-plane couplings are strongly FM, with J$_{Fe1–Fe2}$ and J$_{Fe1–Co}$ exhibiting values of 56.8 and 10.7 meV, respectively. In contrast, the in-plane interactions are AF, where J$_{Fe1–Fe1}$ = -1.3 meV and J$_{Co–Co}$ = -0.8 meV, resulting in a frustrated spin lattice. This behaviour resembles that of Fe$_3$GeTe$_2$, where the strong ferromagnetism arising from inter-plane exchange is diminished due to the AF character of in-plane interactions.[19,51] A direct comparison of the exchange couplings for twisted and untwisted Fe$_2$CoGaTe$_2$ (Figure S9) shows that those are nearly identical, indicating that twisting primarily affects interactions between adjacent layers.

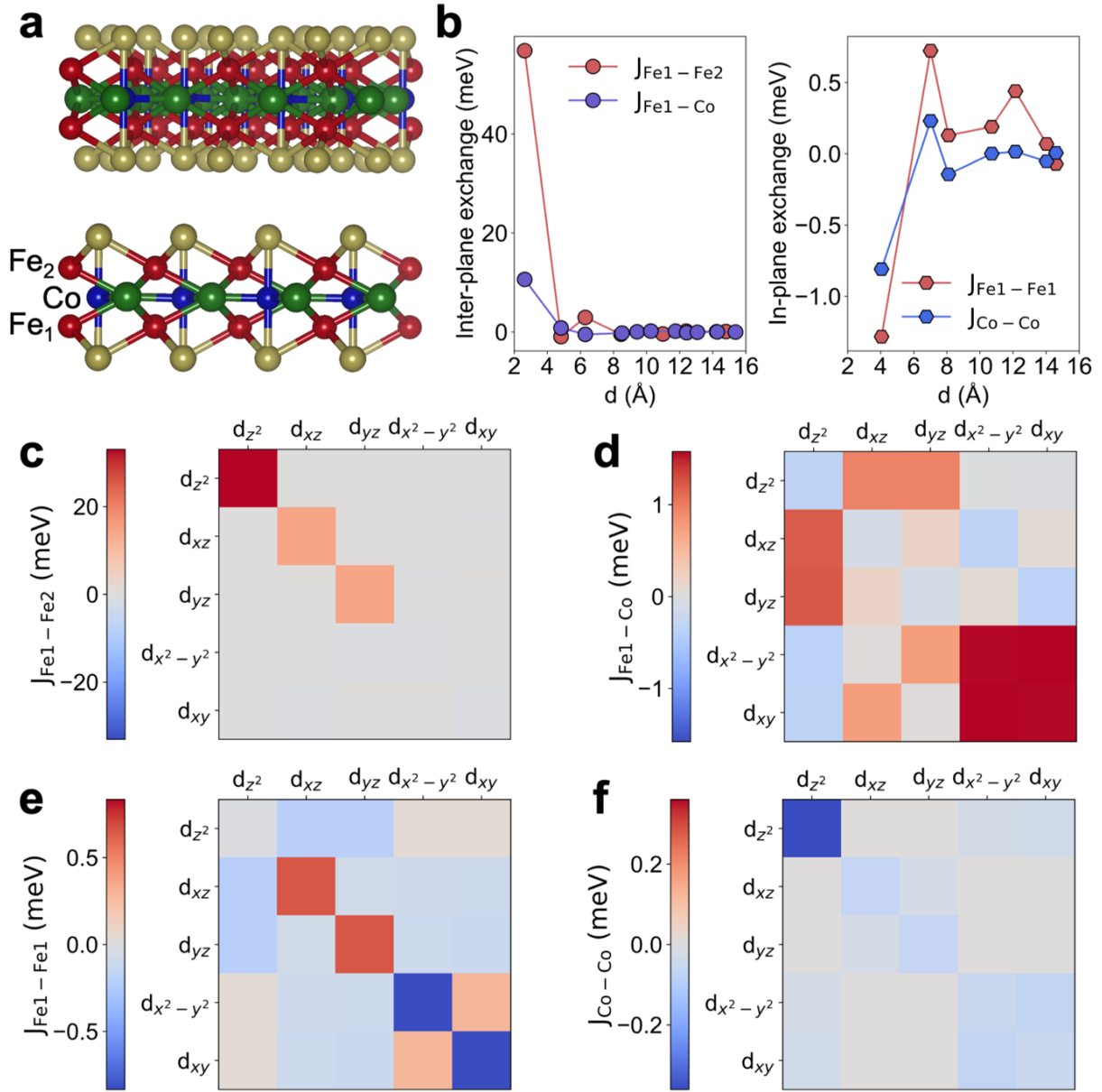

Figure 4. a) Lateral view of twisted bilayer Fe$_2$CoGaTe$_2$ labelling the Fe$_1$, Fe$_2$ and Co atom within one layer. b) Inter-plane exchange interactions J$_{Fe1-Fe2}$, J$_{Fe1-Co}$ (left) and in-plane couplings J$_{Fe1-Fe1}$ and J$_{Co-Co}$ (right) for twisted bilayer Fe$_2$CoGaTe$_2$. c-f) Orbital-resolved contribution to J$_{Fe1-Fe2}$, J$_{Fe1-Co}$, J$_{Fe1-Fe1}$ and J$_{Co-Co}$, respectively.

The contrasting FM and AF character of the inter-plane and in-plane couplings arises from the orbitals stabilizing the long-range magnetic order in this material. This is analysed microscopically through the orbital-resolved contributions to the exchange (Figure 4c-f). We find that the FM character of J$_{Fe1-Fe2}$ is mainly due to the contribution of d$_{z^2}$-d$_{z^2}$, d$_{xz}$-d$_{xz}$ and d$_{yz}$-d$_{yz}$ orbitals, consistent with the behaviour of Fe$_3$GaTe$_2$ (Figure S10 and S11).[19] The enhanced magnitude of J$_{Fe1-Fe2}$ in Fe$_2$CoGaTe$_2$ (56.8 meV) compared to Fe$_3$GaTe$_2$ (35.6 meV) is attributed to the increased magnetic moments of Fe$_1$ and Fe$_2$ atoms (2.59 μ$_B$ and 2.54 μ$_B$ in Fe$_2$CoGaTe$_2$ vs 2.35 μ$_B$ and 2.31 μ$_B$ in Fe$_3$GaTe$_2$). Additionally, the FM nature of J$_{Fe1-Co}$ arises from the contribution of $d_{x2-y2}$-$d_{x2-y2}$, $d_{x2-y2}$-$d_{xy}$ and $d_{xy}$-$d_{xy}$ orbitals, with a more subtle effect of $d_{z2}$-$d_{xz}$ and $d_{z2}$-$d_{yz}$. In contrast, the AF behaviour of $J_{Fe1-Fe1}$ is mediated by the in-plane $d_{x2-y2}$-$d_{x2-y2}$ and $d_{xy}$-$d_{xy}$ superexchange pathways, while antiferromagnetism in $J_{Co-Co}$ mainly originates from a $d_{z2}$−$d_{z2}$ mechanism, mirroring the behaviour of $Fe_3GeTe_2$. This contrasts with $Fe_3GaTe_2$, where the above-room-temperature ferromagnetism is stabilized by the combined FM character of inter-plane and in-plane exchange couplings (Figure S10). We attribute this to the different number of electrons in each material: $Fe_3GeTe_2$ contains one additional electron per Ge atom compared to Ga, and $Fe_2CoGaTe_2$ gains extra electrons from Co substitution relative to $Fe_3GaTe_2$. In both cases, the additional electrons stabilise AF in-plane interactions, leading to a reduction of the overall FM order and thus of the magnetic ordering temperature. Our computed $T_N$ for twisted $Fe_2CoGaTe_2$ is 280 K (Figure S12). This value is expected to be overestimated, as our calculations yield $T_N$ = 290 K for the untwisted bilayer $Fe_2CoGaTe_2$, higher than the experimental value of $T_N \approx 130$ K.[25] Such deviations are common in the determination of $T_N$ and $T_C$ in similar itinerant magnetic compounds.[52,53] Nevertheless, our calculations result in a reduction of the magnetic ordering temperature in $Fe_2CoGaTe_2$ relative to $Fe_3GaTe_2$ by a factor of 2.1, aligning well with experimental findings (Figure S12).[25,39]

Given the interlayer antiferromagnetism and metallic character of Co-doped $Fe_3GaTe_2$, its potential for altermagnetism could be further explored through alternative strategies to realize this state in bulk (untwisted) $Fe_3GaTe_2$-based antiferromagnets. One possible approach is the realization of Janus configurations, where one Te layer is replaced by Se within each layer. Since the bulk unit cell consists of two layers, the resulting symmetry depends on whether the Se substitution occurs on the same or on opposite sides of the top and bottom layers. When the Janus asymmetry is applied on opposite sides of the two layers (i.e., the top side of one layer and the bottom side of the other layer), the system remains centrosymmetric (P-3m1 ($D_{3d}$)) and non-relativistic spin splitting is forbidden. In contrast, if Se atoms are introduced on the same side of both layers, the inversion symmetry is broken, resulting in the non-centrosymmetric $P6_3mc$ ($C_{6v}$) space group, which allows non-relativistic spin splitting at zero net magnetization. Additionally, given that in bulk $Fe_{3-x}GaTe_2$, Fe vacancies induce a symmetry lowering from $P6_3/mmc$ ($D_{6h}$) to the polar P3m1 ($C_{3v}$) space group due to displacement of central Fe atoms,[54–56] a similar mechanism could be exploited via identical out-of-plane displacements of the central magnetic atoms to induce altermagnetism. Such displacements would break inversion symmetry, leading to the $P6_3mc$ ($C_{6v}$) space group and enabling altermagnetism, whereas unequal displacements between central magnetic atoms of top and bottom layer would lower the symmetry to P3m1 ($C_{3v}$), inducing a ferrimagnetic state.

3. Conclusion

In conclusion, symmetry analysis reveals that twisted bilayer $Fe_2CoGaTe_2$ hosts an *i*-wave AM state in the non-relativistic limit. This state arises from the interplay of AF interlayer coupling and the absence of inversion symmetry, thereby breaking the global $\mathcal{PT}$ symmetry. Our first first-principles calculations reveal a pronounced momentum-dependent spin-splitting, reaching a maximum value of 138 meV, alongside with a spin-resolved Fermi surface exhibiting a set of star-like sheets arranged with sixfold symmetry. Upon considering SOC effects, the spin degeneracy is preserved along the directions of the twofold rotation axes, but lifted along the perpendicular directions. Additionally, we provide a deep microscopic analysis of the magnetic exchange interactions governing the $T_N$ in twisted bilayer $Fe_2CoGaTe_2$. We show that the additional electrons introduced by Co atoms drive a FM-to-AF transition of the in-plane exchange couplings relative to pristine $Fe_3GaTe_2$, and thus leading to a lower ordering temperature compared to the latter, mirroring the behaviour of $Fe_3GeTe_2$. Overall, our results position twistronics as a versatile approach to induce altermagnetism in $Fe_2CoGaTe_2$, highlighting the potential of these techniques and related Fe-based metallic van der Waals antiferromagnets for their integration in high-efficiency spintronic devices.

4. Methods

Calculations for bilayer $Fe_3GaTe_2$, $Fe_2CoGaTe_2$ and twisted $Fe_2CoGaTe_2$ were computed by employing the VASP package.[57] We employ the generalized gradient approximation (GGA) to describe the exchange-correlation energy using a plane wave cutoff of 500 eV. Additionally, we employ the LDA functional confirm the robustness of the interlayer AF coupling for twisted $Fe_2CoGaTe_2$. For the simulation of the bilayer structures, we included a vacuum spacing of 15 Å along the vertical direction to avoid spurious interactions with the periodic image. For treating the vdW interactions between layers, the DFT-D2 approximation was employed. Maximally localized Wannier functions were constructed using the Wannier90 package,[58] for which the d orbitals of Fe and Co atoms as well as the p orbitals of Ga and Te were employed for the basis. The exchange couplings were obtained using of the TB2J code[59] and the ordering temperature of the compounds through atomistic simulations as implemented in the VAMPIRE package.[60]

## Acknowledgements

The authors acknowledge the financial support from the European Union (ERC-2021-StG-101042680 2D-SMARTiES), the Spanish Government MCIU (PID2024-162182NAI00 2D-MAGIC) and the Generalitat Valenciana (grant CIDEXG/2023/1). A.M.R. thanks the Spanish MIU (Grant No FPU21/04195). The calculations were performed on the HAWK cluster of the 2D Smart Materials Lab hosted by Servei d'Informàtica of the Universitat de València.